\documentclass[fleqn,usenatbib]{mnras}

\usepackage{newtxtext,newtxmath}

\usepackage[T1]{fontenc}

\DeclareRobustCommand{\VAN}[3]{#2}
\let\VANthebibliography\thebibliography
\def\thebibliography{\DeclareRobustCommand{\VAN}[3]{##3}\VANthebibliography}

\usepackage{graphicx}	
\usepackage{amsmath}	

\title[Closed field line vortices in magnetospheres]{Closed field line vortices in planetary magnetospheres}

\author[Z. Nemeth]{
Zoltan Nemeth,$^{1}$\thanks{E-mail: nemeth.zoltan@wigner.hu}
\\
$^{1}$Wigner Research Centre for Physics, Konkoly-Thege Miklós út 29-33., Budapest H-1121, Hungary\\
}

\date{Accepted XXX. Received YYY; in original form ZZZ}

\pubyear{2022}

\begin{document}
\label{firstpage}
\pagerange{\pageref{firstpage}--\pageref{lastpage}}
\maketitle

\begin{abstract}
In a rotation-dominated magnetosphere, there is a region where closed field lines rotate around the planet, and also a region where the open field lines stretch away from the planet, forming the lobes of the magnetotail. This paper shows that there could be a third, significantly different region, where the closed field lines form twisted vortex structures anchored in the magnetotail. Such patterns form when there are significant plasma sources inside the magnetosphere and the time scale of the plasmoid formation process is substantially larger than the planetary rotation period. In the presence of vortices, the Dungey and Vasyliunas cycles act differently. The Dungey flow does not penetrate the central region of the polar cap. Tail reconnection events are rare, thus leaving the plasma time enough to participate in the essentially 3-dimensional vortex-forming plasma motion. The above conditions are fulfilled for Saturn. We discovered vortex-like patterns in the plasma and magnetic field data measured by the Cassini spacecraft in the nightside magnetosphere of Saturn. The plasma whirling around in these vortices never reaches the dayside, instead, it performs a retrograde motion in the high latitude regions of the magnetotail. Low-energy plasma data suggest that the observed patterns correspond to the closed field line vortices.
\end{abstract}




\section{Introduction}

It is generally accepted that the magnetospheres of giant planets consist of two topologically distinct regions: the region of closed field in the equatorial and mid-latitudes, where both ends of the field lines are connected to the planet; and the region of open field, where the field lines have only one foot-point anchored in the ionosphere, either in the northern or in the southern polar cap. The plasma content of the closed field region is thought to co-rotate with the planet, exhibiting some extent of lag (sub-corotation)  \citep[][]{bunce2003azimuthal, cowley2003corotation-driven, cowley2004simple}. The field and plasma form more-or-less axisymmetric shells in this region, similar to the L-shells of a (quasi-)dipole field configuration. According to the current models, these shells move as rigid bodies: their angular velocity is constant along the field lines from the northern hemisphere through the magnetic equator to the southern hemisphere. The amount of lag (sub-corotation) is a function of the latitude only (or equivalently: a function of the flux-function describing the axisymmetric field \citep[][]{cowley2003corotation-driven, cowley2004simple}, and the gradient of this function determines the nature of the magnetosphere-ionosphere interaction, for example, the position and properties of the auroral ovals \citep[][]{cowley2001origin, bunce2003azimuthal, cowley2003corotation-driven, cowley2004simple}.

The foot-points of the open field lines also (sub)co-rotate with the planet; this motion twists the field lines into a spiral pattern \citep[][]{isbell1984magnetospheric, vasyliunas1983physics}. The mechanism of this field line twist and the formation of the Parker spiral can be described in a common framework, as was shown by \citet{vasyliunas1983physics}. 

Two important properties differentiate giant planet magnetospheres from that of terrestrial planets: their fast rotation rate and the existence of intense plasma sources inside the magnetospheres. Due to their fast rotation, these are entirely rotation-dominated magnetospheres \citep[][]{brice1970ioannidis}; they lack the convection-dominated closed field region found outside the plasmasphere in the terrestrial magnetosphere. The intense plasma sources inflict significant loading on the field lines, which (together with the fast rotation) leads to field line deformation and the formation of a dense plasma sheet near the magnetic equator \citep[see e.g.][]{ gledhill1967magnetosphere, hill1974origin, hill1974configuration, acuna1983physics, persoon2005equatorial, gombosi2009saturn's, arridge2007mass, arridge2008warping, nemeth2011ion}. It also explains the above-mentioned co-rotation lag, since the ionospheric interaction needs to accelerate the new material continuously introduced to the field lines \citep[][]{hill1979inertial, hill1980corotation}. The loading and stretching of field lines also give rise to a so-called planetary wind \citep[][]{ hill1974origin, michel1974centrifugal}, in which the loaded field lines undergo centrifugal instability and move outward from the planet. This is the basis of the Vasyliunas cycle \citep[][]{vasyliunas1983physics}, in which a plasmoid forming reconnection process removes the excess plasma from the loaded closed field lines. The reconnection also shortens the emptied field lines, which then return to the vicinity of the planet thus enabling the cycle to start over.

In the open field region, the Dungey cycle \citep[][]{dungey1961interplanetary} governs the dynamics, in which closed field lines at the dayside magnetopause are opened up and connected to interplanetary magnetic field lines of the solar wind in a reconnection process. These opened-up field lines are convected tailward by the solar wind flow and form the northern and southern open lobes of the magnetospheric tail. A tail-side reconnection and the subsequent motion of the newly closed filed lines towards the dayside close the cycle.

The above-summarized picture of giant planet magnetospheres rests on several implicit assumptions. The first is that the magnetosphere is essentially axisymmetric – meaning that the obvious deviations from axial symmetry do not essentially alter the nature of the flow patterns around the planet. Another assumption is that the flow pattern in the ionosphere alone determines the plasma flow everywhere in the magnetosphere. A third important assumption is that (in a steady state) in every planetary period the mass-loaded closed field lines are emptied by some process, and thus can revert to the essentially co-rotating behavior of empty field lines.

In this paper, we will examine the validity of the above assumptions, and the consequences of deviations from the assumed behavior for the global structure of giant planet magnetospheres. Since these consequences have experimentally verifiable aspects, we compare those with data measured by the Cassini spacecraft while orbiting planet Saturn.

\section{Theory}

Planetary magnetospheres are manifestly not axisymmetric. The vectors of the solar wind velocity and the planetary magnetic dipole represent two nonparallel preferred directions, the existence of which is incompatible with axial symmetry. (In theory, the planetary dipole may lie in the ecliptic plane, but even in such a system the two vectors coincide only at summer and winter solstice.) In other words, the effect of the solar wind deforms the magnetosphere, which manifests as a compression on the dayside and a relative elongation on the nightside. If the nature of the plasma flow in the magnetosphere remains essentially the same as in the axisymmetric case, all the closed field lines still rotate around the planet, although they are somewhat deformed during their round tour – compressed on the dayside and expanded on the nightside. Is this really the only effect of the symmetry breaking on closed field lines? In order to answer this question, we need to investigate the geometry of open and closed field lines in more detail. 

\begin{figure}
	\includegraphics[width=\columnwidth]{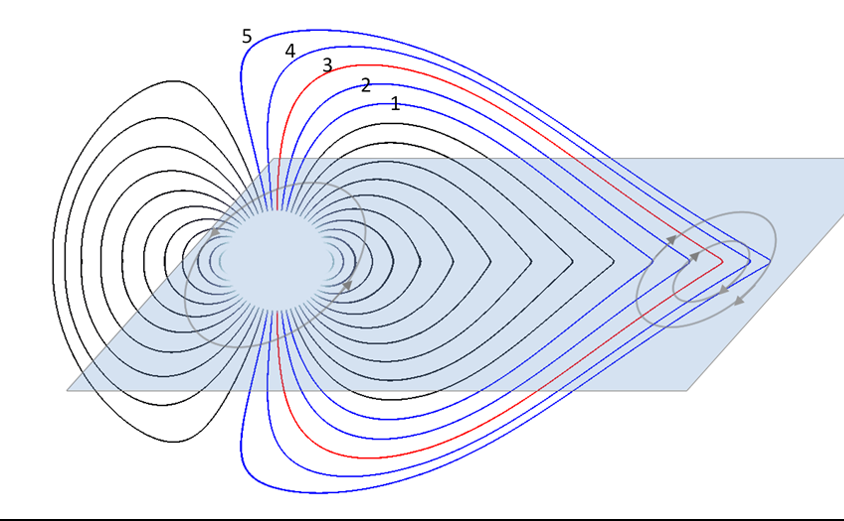}
    \caption{Magnetic field of a dipole with Chapman-Ferraro currents and a tail current sheet. All field lines connected to the planet are closed. Planetary rotation move the field line in position 1 to position 5, and vice versa.}
    \label{fig:1}
\end{figure}

It is generally assumed that those lines which are anchored at the dayside poleward of the cusps, are open field lines. These are the field lines that initially point towards the Sun, but bend back (tailward) after a while. (In stricter terms, the sign of the X component of their tangent vector changes in the polar region of the magnetosphere in a coordinate system in which the X-axis is parallel to the Sun-planet line.) Are they necessarily open field lines? If we start with a simple planetary dipole and disturb it only with Chapman-Ferraro currents, which constrain the planetary field in a cavity inside a perfectly conducting flow, we find that the asymmetry is already present, but all the field lines are closed \citep[][]{mead1964}. Such a configuration can be seen in fig 4 of \citet{mead1964}. There is a critical latitude, above which field lines originating on the dayside pass over the poles and cross the magnetic equator on the nightside. If we add a current sheet (finite in the X direction, very narrow in the direction (Z) of the dipole moment vector, and infinite in the direction (Y) perpendicular to both), we find that the field lines are still closed (Fig.~\ref{fig:1}). 
If we add a homogenous northward field (representing the Interplanetary Magnetic Field (IMF)), the field lines are still closed, and the volume of the magnetosphere is more constrained. Only adding a southward IMF will create open field lines. If we consider the dynamics of this process, we arrive at the Dungey cycle: the southward-directed IMF field lines arriving at the magnetopause reconnect with the closed field lines, and thus open magnetospheric field lines are created. In other words, it is the Dungey cycle, which dynamically creates the open field. Without it (if e.g. the IMF remains northward for a longer time) the magnetosphere is closed, which includes the bent-back field lines originating on the dayside above the critical latitude and closing on the nightside. What happens with these field lines during the rotation of the planet?

Since the footpoints of these field lines are anchored in the ionosphere, these footpoints circle around the planet together with the ionospheric plasma. Although the plasma in and near the polar cap can exhibit significant sub-corotation \citep[][]{stallard2004ion, stallard2019averaged}, it is impossible for the footpoints not to rotate with the planet. However small the pro-grade plasma motion in the ionosphere, it will carry the footpoints around the planet. On the contrary, the middle point of these field lines anchored in the dense equatorial plasma sheet will always remain on the nightside (even when the footpoints lie on the noon meridian). This suggests a plasma motion fundamentally different from the axisymmetric case: since the middle points remain on the nightside while the northern and southern ionospheric footpoints travel around the planet, and the field lines are continuous through this motion, there must be parts of the field line, which (instead of rotating around the planet) sweep over the poles (see Fig.~\ref{fig:1}). This kind of motion, of course, requires that the field line remain continuous during most of a planetary rotation (or more), and thus the plasmoid forming tail reconnection rate should be less than 1 per planetary day. As was shown by \citet{cowley2015down}, this really is the case for Saturn, where the characteristic time between plasmoids is 35-45 h, which is much longer than the 10.5 h planetary rotation period. (Notice that the observed flow violates another of the above-mentioned common assumptions – namely that in every planetary period the mass-loaded closed field lines are disconnected from the planet.)

The planetary rotation will move the field line from position 1 in Fig.~\ref{fig:1}. to position 5, and vice versa. This means that some part of field line 5 (the part close to the middle point, which lies inside and near the equatorial plasma sheet) must first move dawnward, move away from the planet after that, travel duskward in the far tail, then move towards the planet near the dusk side of the magnetopause. It is possible that the middle point cannot complete this circle before a tail reconnection severs the field line, since plasma motion in the far tail can be even slower than that indicated by the angular velocity of the ionospheric footpoints. Still, by and large, it seems that the middle points of closed field lines, which originate from high latitudes of the ionosphere, circle around a point in the far tail, and not around the planet. How is that possible?

First, we should note that the force which balances the centrifugal force that the dense equatorial plasma exerts on the field line, is the $j \times B$ force associated with the sharp bend in the field line inside the current sheet (see Fig.~\ref{fig:1}). When field line 5 moves towards the dayside, the tight horizontal V shape, formed by the loaded field line when its footpoints are near midnight, will open up to be able to move above and below the polar regions. This lowers the curvature force (magnetic tension) exerted by that field line in the plasma sheet region, which upsets the balance of centrifugal and magnetic forces. Thus the plasma will move away from the planet under centrifugal forcing. Similarly, when the footpoints move towards the nightside near dusk, the northern and southern parts of the field line move closer together enhancing the magnetic tension – thus the plasma in the plasma sheet pierced by this field line moves towards the planet. The retrograde motion in the far tail plasma sheet is simply because the Y component of the $j \times B$ force points in the retrograde direction while the footpoints move on the dayside from dawn to dusk.

\begin{figure}
	\includegraphics[width=\columnwidth]{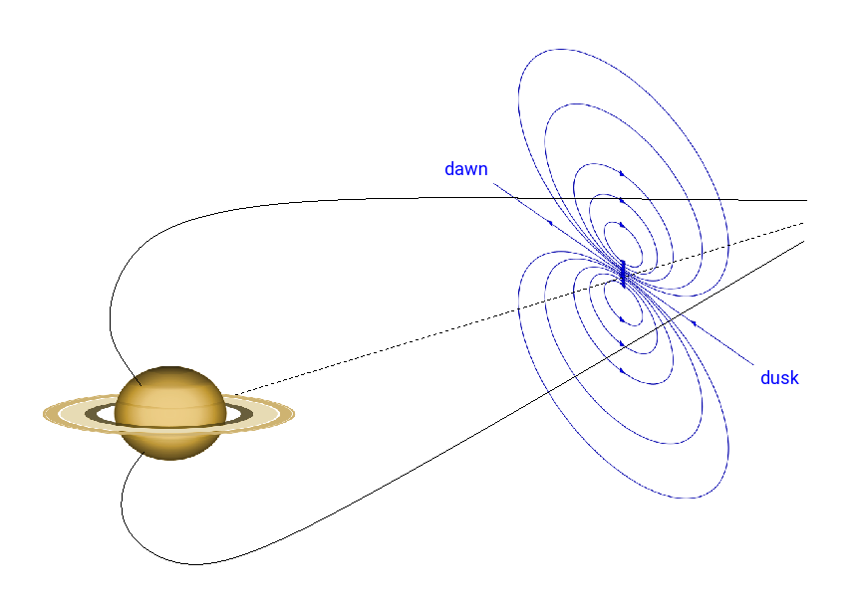}
    \caption{Schematic of the cross-tail plasma flows carrying the field lines in the closed-field vortices.}
    \label{fig:2}
\end{figure}

Although this circular motion in the far tail plasma sheet might not be able to complete in the available time (before reconnection severs the magnetic connection between the footpoints and the far tail plasma sheet), the phenomenon also influences plasma motion in the tail lobes closer to the planet. Field lines that have ionospheric footpoints above the critical latitude and middle points anchored in the far tail, cut out elliptical paths from planes in the middle magnetosphere perpendicular to the X-axis (Fig.~\ref{fig:2}). 
The totality of these field lines forms two giant vortices, one in the northern and one in the southern magnetospheric lobe. The plasma velocities in these two vortices are more or less independent of each other. They approach the same value in the far tail plasma sheet (near the middle points of the field lines), but closer to the planet, in the magnetospheric lobes, the velocities in the northern lobe are determined by the motion patterns of the northern ionosphere, while the southern lobe is governed by the southern ionosphere. Since for a gas giant there could be significant, measurable differences between the velocities of ionospheric flows of mirror latitudes in the northern and southern hemispheres, the characteristic periodicities in the northern and southern magnetospheric lobes can also be different. In contrast, in a quasi-axisymmetric flow pattern, in which the L-shells move as rigid bodies, the periodicities observable in the northern and southern magnetospheric lobes should be the same. Otherwise, the differential rotation of the northern and southern parts of the same L-shell would stretch its field lines indefinitely in the azimuthal direction in the vicinity of the plasma sheet. Observations reporting slightly different periodicities in the two lobes for various magnetospheric phenomena \citep[][]{andrews2008planetary, andrews2010magnetic, andrews2012planetary, provan2012dual, provan2019planetary, provan2021planetary, szego2012location, szego2013dual} suggest that the flow patterns in the two lobes are independent. Another important aspect to be considered about these planetary period oscillations (PPOs) is that the vortex model decouples the observed periodicity and the plasma speed. Since in the outer magnetosphere the plasma is significantly slower than the speed required for rigid corotation, plasma rotating around the planet with this lagging speed would show periodicities much longer than the planetary period but this is not the case. On the other hand, in the vortex model, the plasma rotates not around the planet but around the vortex core. This path is significantly shorter than that going all the way around the planet, and thus it requires much lower speeds to keep up with the planetary periodicity.

Investigating the flow in these hypothetical lobe vortices shows that the plasma exhibits a rapid pro-grade azimuthal motion in the plasma sheet, which is slower and slower as we move away from the magnetic equator towards the vortex core. There is a distance where the azimuthal motion stops outright (in the core of the vortex), and if we move even farther from the magnetic equator, we should observe a retrograde motion of the tenuous plasma of the high-Z lobes. Such a flow pattern should be observable in the plasma measurements. \citet{nemeth2015latitudinal} already identified such a flow pattern in the ion measurements of the Cassini spacecraft, but attributed the decreasing azimuthal velocity to sub-corotation intensifying for larger L values, and did not offer an explanation for the observed retrograde motion. In the next section, we revisit these observations in more detail, extending them to latitudinal as well as azimuthal flow patterns and showing how the experimental data support the existence of giant closed field line vortices in the tail lobes.

\begin{figure}
	\centering
	\includegraphics[width=0.8\columnwidth]{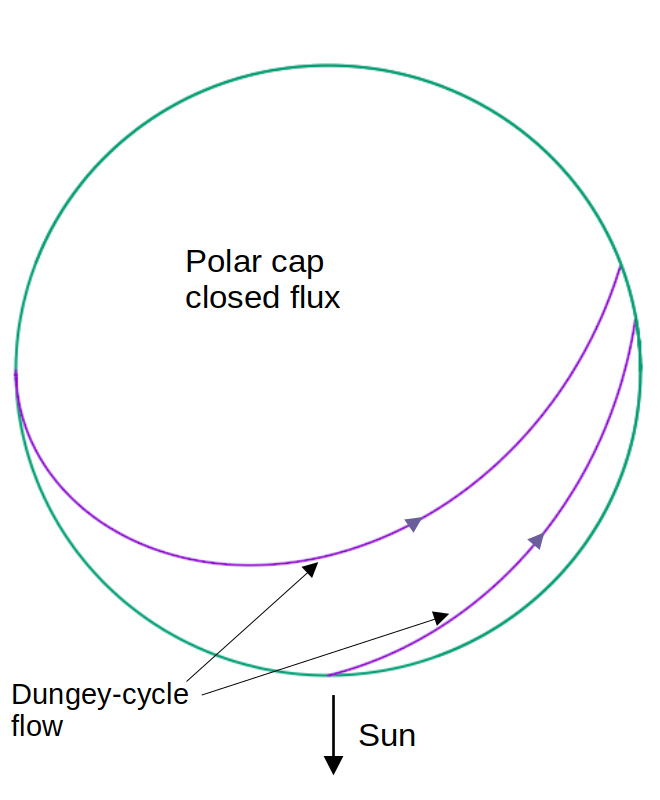}
    \caption{Schematic of the ionospheric plasma flow in the polar cap. Due to the fast rotation, the Dungey-cycle flow cannot penetrate the entire polar cap.}
    \label{fig:3}
\end{figure}

Another aspect of the theory is how the Dungey cycle and the open field lines fit into this picture. If we simply suppose that the open field lines are those closest to the poles (as in the terrestrial magnetosphere), we find that the field lines of the closed field vortices should wind over and around the open field. In other words, the spatially bounded volume of the closed field vortices should encompass the spatially infinite open field lines, which is impossible. To resolve this seeming controversy, we should consider the dynamics of the process: how the Dungey cycle opens up the originally closed field lines of the dayside magnetosphere. At the moment of the dayside reconnection, the footpoints of the reconnecting closed field lines intersect the ionosphere at the critical latitude. Once the reconnection opened up the field line, the convection associated with the Dungey cycle starts to move the footpoint towards the nightside, which at first means a poleward motion. At the same time, the ionospheric plasma rotates around the planet, which adds an azimuthal component to the velocity. \citet{cowley2004simple} estimate the speed of the poleward motion to be 200 m/s in the case of Saturn, while the rotation speed is around 500-700 m/s. Considering these two motions together, we find that the footpoints of the opened-up field line never penetrate the core of the polar cap. Before that could happen, planetary rotation moves the footpoint onto the nightside, where the Dungey cycle flow acts to move the footpoint farther away from the pole. In Fig.~\ref{fig:3} the most extreme case of the footpoint motion is shown, where the field line is opened up at 6 a.m. local time, and thus can penetrate a good portion of the polar cap, but still not all the way to the pole. 
This means that the open field lines lie on (and near) the outer surface of the closed field vortices, it is the open field that winds around the closed field vortex, and not the other way around. Close to the poles reside the undisturbed cores of the open field vortices. This may be related to the decreased corotation lag observed near the poles \citep[][]{stallard2019averaged}.

Thus the following picture describes the magnetospheric structure of giant planets, provided that they are fast rotators, there are significant plasma sources inside the magnetosphere, and the characteristic time, during which the far tail plasma sheet remains connected to the planet, is longer than the planetary period: Close to the planet, at equatorial and mid-latitudes, there still is a region where the plasma bound to closed field lines rotate around the planet. At a critical latitude, depending on solar wind conditions (most notably on the orientation of the IMF) the field line topology changes (open-closed field boundary). Near this latitude, the footpoints of open field lines circle around the planet. On even higher latitudes we again find closed field lines; here the giant closed field line vortices connect to the planet. The middle points of field lines in these vortices are anchored in the far tail plasma sheet. The rotation of the plasma in the vortices forms a distinctive velocity pattern in the tail lobes, characterized by retrograde plasma motion far away from the magnetic equator. The open field lines wind around the closed field vortices. For periods of rapid plasmoid formation and strong southward-directed IMF, the vortices may disappear, in which case the classic picture describes the magnetospheric structure. Although it is difficult to judge the global structure (magnetic connectedness) from local measurements, it will be shown in the next section that, in the case of Saturn, the measured flow patterns and field directions support the magnetic vortex picture.

\section{Data}

Saturn is a fast-rotating gas giant with an extended magnetosphere. The Pioneer and Voyager probes performed the first in situ measurements in the Kronian magnetosphere, as they flew by the planet in 1979, 1980, and 1981. Further data were provided by the Cassini orbiter between 2004 and 2017. The analysis of these measurements revealed the unique and complex structure of Saturn's magnetosphere, the results of which are summarized in several review studies \citep[][]{gombosi2009saturn's, mitchell2009dynamics, mauk2009fundamental}. 

In this section, we investigate in situ data measured by the Cassini spacecraft in the nightside outer magnetosphere of Saturn, including magnetic field data provided by the Cassini Magnetometer \citep[][]{dougherty2004cassini} (MAG)  and the azimuthal and latitudinal components of the plasma velocities (H$^+$ and water group ions) from the LANMOM numerical ion moments derived by \citet{thomsen2010survey} from the measurements of the Cassini Plasma Spectrometer (CAPS) \citep[][]{young2004cassini}. 

Our analysis is based on data from 2006 and 2009 in the southern summer period, as the spacecraft in these 2 years spent a significant amount of time exploring the nightside outer magnetosphere of Saturn. We analyze orbit segments containing Titan encounters because these segments provide the best latitudinal scans of the tail region together with relatively small radial motion. We use data in which the Kronian local time (LT) is less than 3 hours from midnight and where the distance of Cassini from Saturn is 20±4 Saturn radii ($R_S$).

\begin{figure}
	\includegraphics[width=\columnwidth]{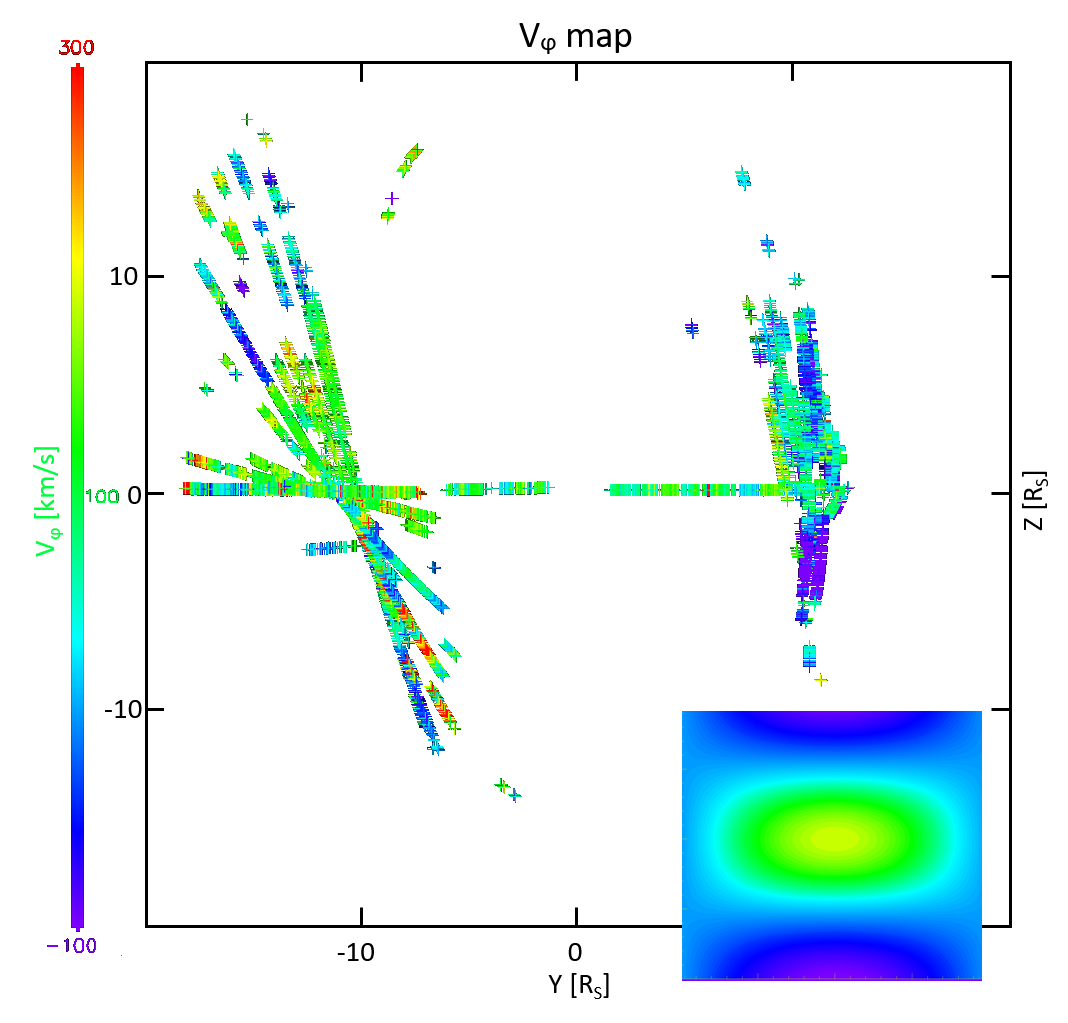}
    \caption{Cross-tail map of the azimuthal plasma speed. The inset shows the pattern expected in a vortex pattern.}
    \label{fig:4}
\end{figure}
Fig.~\ref{fig:4} shows a cross-tail map of the azimuthal plasma speed, projected onto a plane perpendicular to the Sun-Saturn direction and crossing the tail at the position of Titan. Cold colors (dark blue and purple) represent retrograde plasma motion. The inset shows the velocity pattern expected if two vortices (one in each tail lobe) determine the plasma motion. We do not expect one-to-one correspondence, since the data set covering two complete Earth years suffers from significant time variability (the most prominent of which is the “flapping” of the Kronian magnetodisk \citep[][]{simon2010titan's, arridge2011periodic, szego2012location, szego2013dual}. Despite this, the overall resemblance is quite prominent, the model describes the measurements much better than the rotating shell picture, in which the map would feature continuous stripes parallel to the equator and no retrograde motion at all.

\begin{figure}
	\includegraphics[width=\columnwidth]{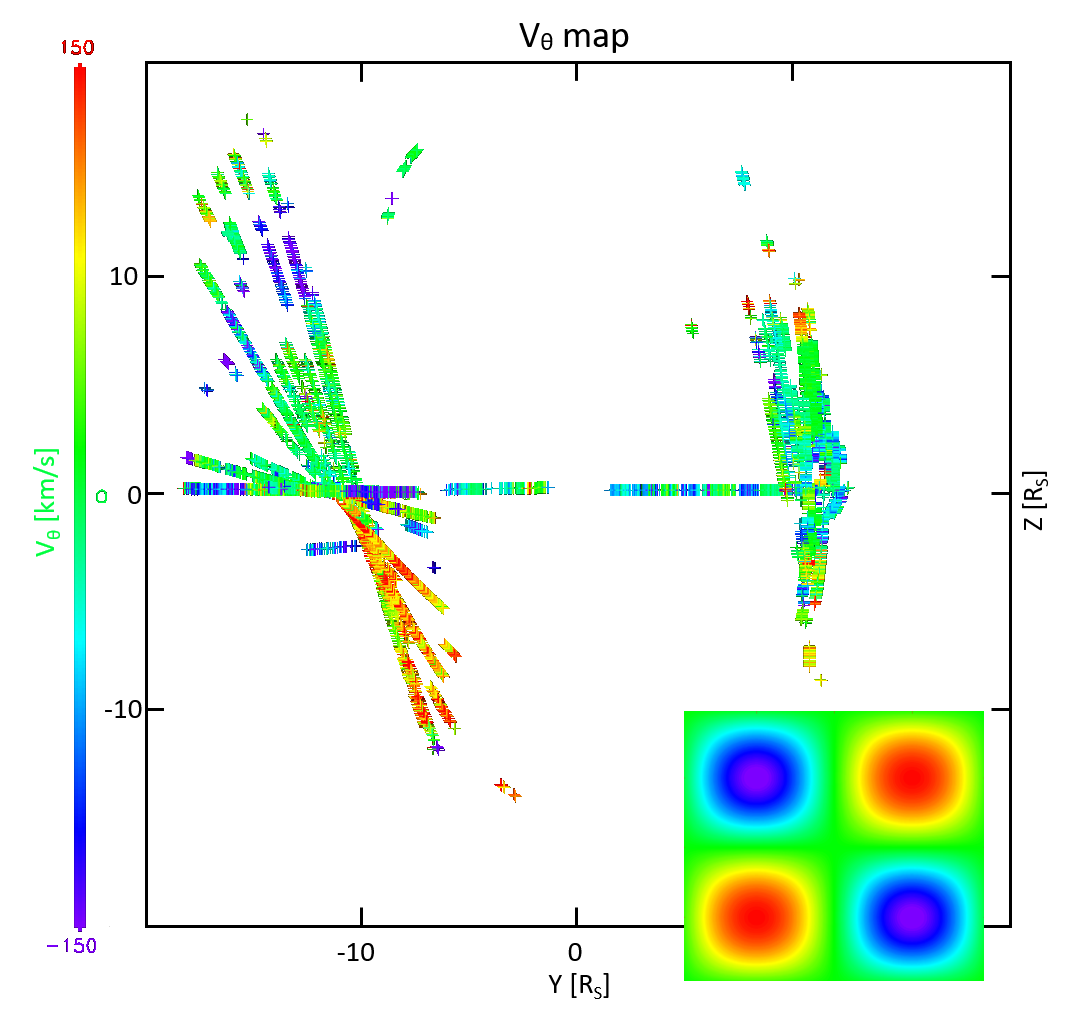}
    \caption{Cross-tail map of the latitudinal plasma speed. The inset shows the pattern expected in a vortex pattern.}
    \label{fig:5}
\end{figure}
Fig.~\ref{fig:5} shows a similar map of the latitudinal plasma speed, with the expected velocity pattern shown in the inset. It is clearly evident that the plasma moves away from the equatorial plane on the dawn side of the tail in accordance with our expectations. We expect plasma motion towards the equatorial plane on the dusk side of the tail. This is not so readily apparent in the measurements, although the data is compatible with this notion as well.

It is also apparent from both figures that the center of the plasma sheet is offset towards the northern lobe. This is a consequence of the magnetodisk being deformed by solar wind loading as shown by \citet{arridge2008warping}.

The last supporting evidence is the distribution of the magnetic field direction during this time period. The root cause of the magnetospheric plasma motion in the rotation-dominated magnetospheres of giant planets is the force that the ionosphere exerts on the plasma through the magnetic field. In other words, the ionospheric motion drags the magnetospheric plasma by means of field-line tension. Since the bulk of the plasma resides in the plasma sheet, the force density – and thus the corresponding field line curvature – is the largest there. Evidently, the sign of the radial magnetic field reverses inside the current sheet, but the curvature force corresponding to this directional variation is not related to azimuthal forcing, it balances the centrifugal force. The force which accelerates the plasma in the azimuthal direction corresponds to a direction change of the azimuthal component of the magnetic field. For rigid corotation, the field lines lie inside a radial-latitudinal plane. If the plasma of the current sheet lags behind the ionospheric plasma, the magnetic field direction outside the magnetic equator deviates from the radial-latitudinal plane. Near the magnetic equator, the plasma is dragged in the pro-grade direction, which means that the field line deviation is also pro-grade with respect to the radial-latitudinal plane. At the magnetic equator, the field has only a latitudinal component. As a first approximation, we expect a linear increase of the azimuthal field component with the distance from the magnetic equator. That must be true for all models in the vicinity of the magnetic equator, but the model predictions deviate for larger distances. For co-rotating shells, there is a monotonic, although diminishing increase with distance, since the ionosphere always precedes the plasma sheet in the pro-grade direction. For the closed field line vortex model, after reaching a maximum, the azimuthal component starts to decrease. At the center of the vortex, where we encounter a field line that magnetically connects the plasma sheet to the noon meridian, the azimuthal field component is zero.  Farther away from the magnetic equator, in the region of retrograde plasma motion, the field lines deviate from the radial-latitudinal plane in the retrograde direction, thus the azimuthal component reverses there. 
\begin{figure}
	\includegraphics[width=\columnwidth]{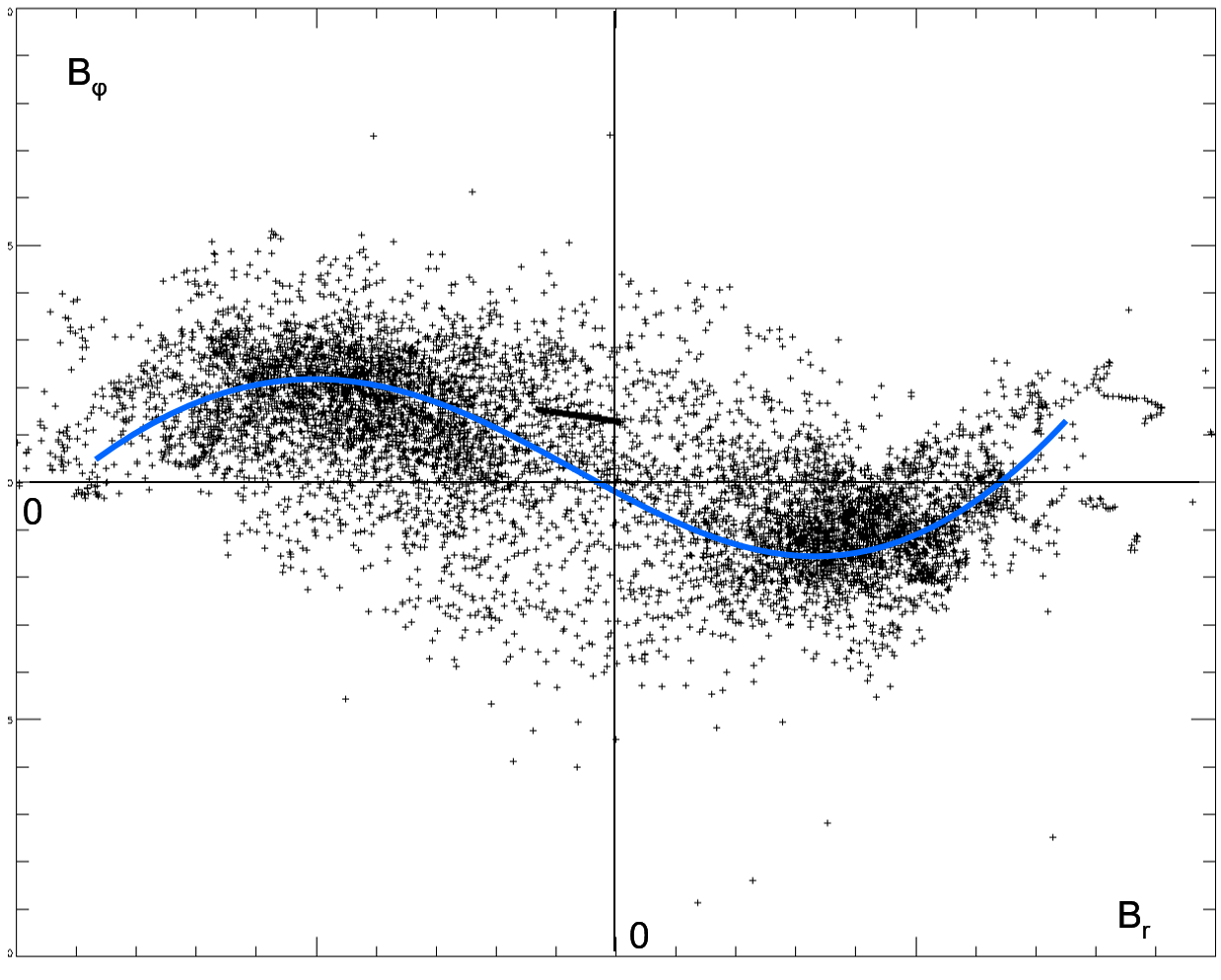}
    \caption{Azimuthal magnetic field $B_\varphi$ as a function of the residual radial magnetic field $B_r$. As one moves away from the magnetic equator, $B_r$ increases monotonically, $B_\varphi$ has a maximum and turns around in a vortex pattern.}
    \label{fig:6}
\end{figure}
In Fig.~\ref{fig:6} we show the azimuthal magnetic field as a function of the radial magnetic field component, the latter representing the distance from the magnetic equator (since it is a monotonic function of said distance, and we can eliminate the effects of magnetodisk flapping this way \citep[see][]{nemeth2015latitudinal, nemeth2016periodic}. We can see a sinusoidal field behavior, which agrees with the closed field line vortex model – linear increase, decrease, zero, and a directional change as one moves away from the magnetic equator.

It is an important and difficult question whether the plasma and field measurements discussed in this section correspond to closed field lines, or whether the measurements show the whirling of plasma in open field lobes – similar to the phenomenon first detected by \citet{isbell1984magnetospheric} in the Terrestrial magnetosphere. In-situ magnetic field measurements cannot provide conclusive answers about the global topology of a field line. The plasma content provides important clues about connectedness, although it still remains a difficult question, even for Earth \citep[][]{chisham2004whose}, where a multitude of spacecraft provide a wealth of relevant data. On closed field lines, one expects to find the plasma of the internal magnetospheric sources; open field lines, on the other hand, should be depleted of magnetospheric plasma but may contain solar wind particles. The field lines discussed in our analysis contain a significant amount of magnetospheric plasma, even those, characterized by retrograde motion, carry heavy water-group ions. The plasma content at higher latitudes is more dilute than that of the central plasma sheet, but an exponential decrease is expected due to centrifugal confinement \citep[][]{sergis2011dynamics, persoon2020et}. The plasma density changes smoothly in the region in question, in accordance with an initially exponential fall-off. There are no detectable boundaries where the magnetospheric plasma abruptly disappears \citep[see][]{nemeth2015latitudinal}. Thus the behavior of the thermal plasma supports the notion that these are indeed closed-field vortices.

If one examines, however, the hot electron population \citep[see e.g.][]{bunce2008origin}, it turns out that the high latitude field lines are empty of those few hundred eV electrons, which are present on lower latitudes. Due to their low mass and high energy, these hot electrons cannot be confined centrifugally – thus several authors interpret their absence as proof that these field lines are open. One possibility to resolve this conflict would be to take into account the effects of the polarization (ambipolar) electric field. In such a scenario, the centrifugally confined cold heavy ions exert an electric force on the hot electron component, achieving their confinement in the vicinity of the plasma sheet. One can even argue that the measurements show a deceleration of the hot electrons as one moves away from the equatorial regions – their energy distribution shifts towards lower energies as expected from an electrically confined particle population \citep[see e.g. the first panel of fig 4 in][]{bunce2008origin}. Unfortunately, models computing the magnitude of the ambipolar electric potential in the Kronian magnetosphere report more than one order of magnitude lower values than that necessary to confine the hot electrons. (\citet{maurice1997magnetic-field-aligned} report 30V, and \citet{persoon2020et} report 10-20V). It is outside of the scope of this paper to discuss numerical plasma models of the Kronian magnetosphere, but we should note that \citet{maurice1997magnetic-field-aligned} find that the electric field can be as high as 80 V/$R_S$ if there is a cold oxygen component present, but they left out this possibility from their final simulation because the pre-Cassini state of the art \citep[][]{richardson1995extended} did not know about cold water-group ions in the outer magnetosphere. Since there is a cold water component, and the field should be integrated over tens of $R_S$ there, it is entirely possible that a more accurate simulation would result in a potential drop of several hundred Volts. The simulations of \citet{persoon2020et}, on the other hand, focused entirely on the thermal plasma, their initial assumptions left out the hot electron component, which is one of the crucial ingredients to having a sizable ambipolar field. Thus the possibility of electric confinement of hot electrons cannot be ruled out.

In summary, the evidence coming from the hot electrons and that coming from cold plasma seem to contradict each other, but some effects can dampen the hot electron population on the high latitude regions of closed field lines. Thus, based on the evidence of a significant amount of water-group ions being present on these field lines and moving according to the vortex pattern (even performing retrograde motion), the observed vortices are most probably composed of closed field lines.

\section{Conclusions}

We have presented theoretical considerations showing that the coexistence of several conditions eventuates the presence of giant closed field line vortices in planetary magnetospheres. The first two conditions are the presence of significant plasma sources inside the magnetosphere and the planet being a fast rotator. Together these conditions ensure that the magnetosphere possesses a centrifugally forced dense equatorial plasma sheet. A third (although not completely independent) condition is that the time scale of the periodic process, responsible for emptying the mass-loaded closed field lines, is substantially larger than the planetary rotation period, and thus the tail field lines remain connected to the planet during (at least) a full rotation. The fourth and last condition, which is always satisfied for the solar-wind-loaded magnetospheres of the solar system, is that some effect breaks the axial symmetry of the system, introducing a strong day-night asymmetry, and creating field lines, which connect the dayside ionosphere to the nightside plasma sheet. These conditions ensure that the ionospheric footpoints of certain field lines perform (at least) a full rotation around the planet, while their middle (equatorial) point is anchored in the sub-corotating nightside plasma sheet. Instead of rotating around the planet, the plasma trapped on these field lines rotates around a vortex line, which connects the pole with a point in the nightside plasma sheet, thus forming two vortices, one in each tail lobes. The motion patterns of these two vortices are more or less independent, they are only connected in the far tail equatorial region. Thus the periodicities of the plasma motion in the two lobes are independent of each other as well. This allows the plasma properties in the nightside magnetosphere to have dual periodicities, both close to the planetary period.

It turns out that the magnetosphere of Saturn satisfies the above-mentioned conditions, and thus it is a valid question whether the Kronian magnetosphere features giant closed field line vortices. Careful examination of the plasma velocities and the magnetic field direction reveals vortex patterns in the nightside outer magnetosphere of Saturn. The supporting evidence includes retrograde plasma motion far from the magnetic equator, flow towards and away from the plasma sheet on the dusk- and dawnside respectively, independent periodicities in the northern and southern lobes, and the field line geometry showing vortex-like characteristics. 

The newly discovered vortex pattern is either evidence of open-field vortices (similar to that observed in the Terrestrial magnetotail \citep[][]{isbell1984magnetospheric}), or that of the closed field line vortices. Based on thermal plasma measurements, we argue that there are closed-field vortices in the Kronian magnetosphere.

In the vortex model, the Dungey and Vasyliunas cycles act somewhat differently. The Dungey flow does not penetrate the central region of the polar cap due to the fast rotation of the ionosphere. Thus the open field lines reside in the outer layer of the vortices. The plasmoid-forming tail reconnection events necessary to close the Vasyliunas cycle are rare, thus leaving the plasma time enough to participate in the essentially 3-dimensional vortex-forming plasma motion.

\section*{Acknowledgements}

The author would like to thank Stan Cowley for the helpful discussions. This work was supported by the ÚNKP-18-4 New National Excellence Program of the Ministry of Human Capacities and the János Bolyai Research Scholarship of the Hungarian Academy of Sciences. 

\section*{Data Availability}

Calibrated magnetic field data and Cassini ion moments from the Cassini mission are available from the NASA Planetary Data System (https://pds.nasa.gov/).



\bibliographystyle{mnras}
\bibliography{Magvortices_v2} 








\bsp	
\label{lastpage}
\end{document}